\documentclass[12pt,preprint]{aastex}
\begin{document}
\title{
 Constraints Imposed by a High Magnetic Field on Models for the EUV Emission
in the Coma Cluster}
\author{Ming Yann Tsay and Chorng-Yuan Hwang}
\affil{Graduate Institute of Astronomy, National Central University, Taiwan}
\email{hwangcy@astro.ncu.edu.tw}
\and
\author{Stuart Bowyer}
\affil{Space Sciences Laboratory, University of California, Berkeley, CA 94720-7450}
%------------------------------------------------
%------------------------------------------------
\begin{abstract}
  A variety of models have been explored in regard to the origin of the
excess extreme ultraviolet ($\sim 0.1$ keV) emission in the Coma cluster.
It is now established that the flux is non-thermal and the only non-thermal
source mechanism that appears viable is inverse Compton emission produced
by $\sim 100$ MeV electrons interacting with the cosmic microwave background
photons. All but one of the models that have been proposed require a cluster
magnetic field $< 1\mu$G. However, recent observations strongly suggest the
magnetic field in the Coma cluster is $\sim 5 \mu$G. We investigate the
constraints on models imposed by a $5\mu$G cluster field and find a limited
class of models that are compatible with this constraint. We also investigate
the possibility that the excess hard (40-60 keV) X-ray emission in the cluster
is produced by inverse Compton emission with the same electron population that
produces the EUV excess. 
We find no scenarios that are compatible with a
large cluster magnetic field, and consequently in this case these two
components must be unrelated.

\end{abstract}
%----------------------------------------------

\newpage
\section{INTRODUCTION}
  EUV emission in the Coma cluster in excess of that produced by the thermal
X-ray gas was first reported by Lieu et al. (1996). Further analysis of data on
this cluster was carried out by Bowyer and Bergh\"ofer (1998) and Bowyer et al.
(1999). Initially, this flux was attributed to thermal emission from "warm" gas
 at $10^6$ K. Maintenance of a warm intracluster gas is extraordinarily
difficult, and on these grounds alone it was generally agreed that a thermal
source was untenable. Observations relevant to this issue were obtained with
the Hopkins Ultraviolet telescope (Dixon et al. 1996), and FUSE (Dixon et al.
 2001a,b). No Far UV line emission from gas at $10^6$ K was detected. More
recently, observations of several clusters with XMM (Tamura et al. 2001; Kaastra
et al. 2001; Peterson et al. 2001) have shown no evidence for a $10^6$
temperature gas. All additional clusters examined with XMM also show no
evidence for a ``warm'' $10^6$ gas (S. Kahn, private communication). The sum of
 all these findings seems compelling: a thermal mechanism for the EUV excess
can be ruled out. (However, for an alternate point of view see Mittaz et al.
1998; Lieu, Bonamente, \& Mittaz 1999; Lieu et al. 1999; Lieu, Bonamente, \&
Mittaz 2000; and Bonamente, Lieu, \& Mittaz 2001a,b.)

  Since the source mechanism is not thermal, it must be the product of a
non-thermal process. Inverse Compton scattering (ICS) of cosmic rays with the
 2.7 K cosmic microwave background was suggested early-on as a possible source
mechanism (Hwang 1997; En$\beta$lin \& Biermann 1998). This is the only
non-thermal mechanism that has been suggested as a possibility for the source
of this flux.

  A number of researchers have explored models of ICS in an attempt to explain
the EUV excess in the Coma cluster (Hwang 1997, Bowyer \& Bergh\"ofer 1998,
En$\beta$lin \& Biermann 1998, Sarazin 1999, Atoyan \& V\"olk 2000, Brunetti
et al. 2001). With one exception (Atoyan \& V\"olk 2000), all of these models
require a low ($< 1\mu$G) field and typically these models require a very low
magnetic field ($<0.5\mu$G) field.

  A hard (25-80 keV) X-ray flux in excess of that produced by the $10^8$ K
thermal X-ray gas has been detected with observations with $BeppoSAX$
(Fusco-Femiano et al. 1999). Brunetti et al. (2001) have attempted to explain
both the EUV and hard X-ray excesses in Coma by a single complex model. This
model also employed a low magnetic field.

  New studies of the magnetic field in clusters of galaxies strongly suggest
that magnetic fields in clusters of galaxies are $\sim 5\mu$G. In this paper we
 investigate constraints imposed by a high magnetic field on models capable of
producing the EUV excess in clusters of galaxies via the ICS process. We also
briefly consider the effects of a high magnetic field on attempts to produce
both the EUV and the hard X-ray emission by the
ICS mechanism with the same population of cosmic ray electrons.

\section{ASSUMPTIONS AND OVERALL APPROACH}
  We make the following assumptions.

  A.) The magnetic field is $5 \mu$G. The magnetic field in the Coma
cluster has been the subject of considerable study and observational results
vary considerably. Estimates based in part on equipartition arguments typically
result in low field estimates (Rephaeli 1988, Giovannini et al. 1993). However,
recent observational results strongly indicate that cluster magnetic fields are
 quite large. Clarke et al. (2001) studied sixteen clusters; a substantial
number of data points were obtained for each cluster. They found the cluster
magnetic fields were all in the range $\sim 4-7\mu$G. These results seem
compelling, but it could be argued that they are not applicable to the Coma
cluster since that cluster was not included in their study. However, the Coma
cluster field can be determined by a different procedure albeit for only one
location in the cluster. Feretti et al. (1995) studied the rotation measure of
the cluster-embedded head-tail radio source NGC 4869 (5C4.81). They found the
number of magnetic field reversals through the cluster $> 200$ indicating B
$> 4.9\mu$G. This measurement is limited by the resolution of the VLA
configuration employed and, in principle, is a lower limit for the number of
magnetic field reversals through the cluster. It is possible that with even
higher resolution the number of magnetic field reversals through the cluster
could be larger, with a resulting increase in the estimate of B. However, this
estimate is within the 4 to 7 $\mu$G range found by Clarke et al. (2001) for all
 clusters in their sample. In view of these observations, we find the
assumption of 5$\mu$G for the field in the Coma cluster to be quite reasonable.

  B.) A scattering process is operative in the cluster which will result in a
sufficiently long path length for the electrons that the deposition of the
energy of the electrons occurs within the cluster itself. We note there is
overwhelming observational support for this assumption since if the cosmic rays
 did not deposit their energy within the cluster, they would escape the cluster
 and produce excess EUV emission over a region which is far larger than
 the cluster itself.

  C.) We assume the evolution of the electron population is given by the
equation (Sarazin 1999):
\begin{equation}
   \frac{\partial N(\gamma )}{\partial t}= \frac{\partial [b(\gamma )N(\gamma)]
}{\partial \gamma}+Q( \gamma)
\end{equation}
 where $N(\gamma )d \gamma $ is the total number of electrons in the range
$\gamma$ to $\gamma + d \gamma$, $b(\gamma )$ is the energy loss rate of an
 electron with energy of $\gamma$, $Q(\gamma )d\gamma $ is the injection
rate of electrons in the energy range  $\gamma$ to $\gamma + d \gamma$.

  D.) We assume the injected cosmic-ray electrons follow a power law
distribution.

  E.) We assume the only energy loss mechanisms are inverse Compton emission,
synchrotron radiation, Coulomb ionization and bremsstrahlung.

  The energy loss rate for relativistic electrons interacting with the cosmic
microwave background is given by Longair (1994):
\begin{equation}
   b_{IC}( \gamma )= \frac{4}{3} \frac{{\sigma}_T}{m_e c} {\gamma}^2 U_{CMB} =
1.37 \times 10^{-20} {\gamma}^2 {\rm s}^{-1}
\end{equation}
where ${\sigma}_T$ is the Thomson cross section and $U_{CMB}$ is the energy
density of the cosmic microwave background. The loss rate by synchrotron
emission is given by Longair (1994):
\begin{equation}
   b_{syn}(\gamma) =\frac{4}{3} \frac{{\sigma}_T}{m_e c} {\gamma}^2 U_B = 1.30
 \times 10^{-21} {\gamma}^2 ( \frac{B}{1 \mu G})^{2} {\rm s}^{-1}
\end{equation}
 where $B$ is the magnetic field and $U_B = {B^2}/8\pi$ is the energy
density of the magnetic field. The loss rate for Coulomb ionization is given by
 Sarazin (1999):
\begin{equation}
    b_{Coul}(\gamma ) \sim 1.2 \times 10^{-12} n_e \,[1.0 + \frac{\ln
(\gamma / n_e)}{75}]\, {\rm s}^{-1}
\end{equation}
 where $n_e$ is the thermal electron density in the ICM of the Coma cluster.
The loss rate for the Bremsstrahlung is approximately (Sarazin 1999):
\begin{equation}
    b_{brem}(\gamma ) \sim 1.51 \times 10^{-16} n_e \gamma [ \ln (\gamma )
+0.36] {\rm s}^{-1}
\end{equation}

  A comparison of the energy loss rates for these four mechanisms is given in
Fig.1. This shows that ICS is the dominant energy loss rate mechanism for
relativistic electrons in a $0.2\mu$G cluster field while synchrotron radiation
 is the main energy loss rate process for relativistic electrons in $5\mu$G
cluster field. A diagram of the energy loss rate timescale 
${\gamma}/(b_{IC}+b_{syn}+b_{Coul}+b_{brem})$
versus electron energies is shown in Fig.2.
 Note that there is a decrease of 10$\%$ in the energy of the peak energy loss,
 and, more importantly, there is a decrease in the peak energy loss rate of
$\sim 40 \%$ for the case of a $5\mu$G field as compared to a $0.2\mu$G field.

  The observed EUV ICS flux is given by Blumenthal \& Gould (1970):
\begin{equation}
  f_{IC} (\epsilon )= \frac{1}{4 \pi D^2} \times \frac{r_o^2}{{\hbar}^3 c^2
\pi} \times K_e \times (kT)^{\frac  {p+5}{2}} \times F(p) \times {\epsilon}^
{-\frac{p+1}{2}},
\end{equation}
where the symbols have their standard meaning. The energy of the photon
observed, $\epsilon$, has been obtained by folding the effective area of the
EUVE Deep Survey Telescope (Sirk et al. 1997) with absorption of EUV emission by
 the Galactic interstellar medium in the direction of Coma following the
prescription of Bowyer et al. (1999). The peak of the resultant telescope
effective area is at 80 \AA.

  The synchrotron radio flux is given by Blumenthal and Gould (1970):
\begin{equation}
  f_{syn}(\nu ) = \frac{1}{4 \pi D^2} \times \frac{4\pi e^3}{m_e c^2} \times
K_e \times B^{\frac{p+1}{2}} \times (\frac{3e}{4\pi m_e c})^{\frac{p-1}{2}}
\times a(p) \times {\nu}^{\frac{-p+1}{2}}.
\end{equation}
We have used an integrated diffuse flux density for Coma C of 640 mJy at 1.4
GHz (Deiss et al. 1997) in combination with a spectral index $\alpha = (p-1)/2
= 1.16$ (Bowyer \& Bergh\"ofer 1998) as our observational base.

  We investigate three scenarios. In the first, we consider a large injection
event occurring over a relatively short time which produces the cosmic-ray
electron population. In the second scenario, we assume a continuous sequence of
 small injection events. Finally, we consider a combination of these two: a
large injection event followed by continuous smaller events. In Table 1 we show
 the input parameters for these three models.

  In this Table, $N_o$ is the normalization factor and $p$ is the spectral
index of the initial injected rate. $Q_o$ refers to the normalization factor
and $p'$ is the spectral index of the continuous injection rate. $Q_o$ is
related to the injected  electron number $Q$ by the relation $Q= Q_o \times 
{\gamma}^{-p'}$.
$n_e$ is the thermal electron density in the ICM. EI denotes the
energy input.

\section{RESULTS}
  We first consider the case of a single primary injection event. The results
with input parameters from  Model 1 are shown in Fig.3. This population will 
produce the EUV emission as reported by Bowyer et al. (1999) shown as a cross. 
This population greatly overproduces the observed synchrotron radio emission 
which is shown as a heavy solid line in the figure. With the passage of time, 
synchrotron losses degrade the higher energy electrons, and the resultant 
population after $1.4\times 10^8$ yr and after $2.5 \times 10^8$ yr 
is shown by line 
A and by line B respectively. The results shown in this figure demonstrate why 
it is not possible to produce both the EUV and the radio emission by a 
population of electrons produced in a single initial event.

  We next consider the case of a continuous distribution of smaller events.
The results with the input parameters from Model 2 are shown in Fig.4. The
dashed line shows the total injected electrons. The thin solid line shows the
electron population after $2\times 10^9$ yr. The heavy solid line shows the
distribution of the electrons that are required to produce the observed
synchrotron radio emission and the cross shows the distribution of the
electrons producing the observed EUV emission. The results shown in this figure
 show why no uniform continuous distribution of injection events can reproduce
the observational data.

  We have independently derived a variety of electron populations using a
cluster field of $0.2\mu$G; these are all capable of reproducing the EUV and
radio observations. This is consistent with results found by a number of
previous studies that demonstrated that a continuous distribution of injection
events can produce populations that will replicate both the EUV and the radio
data in the case of low cluster fields.

  We next consider a primary event followed by continuous injections. We use
input parameters listed in Models 3 through 11. The population obtained using
the parameters in Model 3 after an evolution time of 1 Gyr are shown as a
dashed line in Fig.5. We also show the effect of changes in the size of the
initial event. These results are shown as line A and employ inputs from Model
4. Finally, we show the effect of changes in $p$. Using inputs taken from Model
 5, we obtain the population shown as line B.

  In Fig.6 we show the effect of changes in $Q_o$ and in the spectral index of
the electrons in the input events, $p'$, on the overall electron population.
The inputs for the dashed line are from Model 3. Line A shows the results of
decreasing the spectral index using inputs from Model 6. Line B shows the
results of decreasing $Q_o$ using inputs of Model 7. The results in this figure
 show that allowed variations in $Q_o$ and $p'$ are quite small since even
modest changes in one of these parameters alone will result in an electron
population that cannot produce the radio data.

  We have investigated the extent to which we can mimic the best fit electron
population shown in Model 3 by  models which change $p'$ and $Q_o$ (the
continuous injections). Inputs are shown in Model 8 and 9. The resultant
electron distributions are shown in Fig.7. As can be seen there is only a
limited amount of freedom in constructing this type of model since large
changes in these offsetting parameters which can be used to fit the EUV data
will quickly destroy the fit to the radio data. Good fits will be obtained if
$p'$ is $\sim$ 2.32.

  In Fig.8 we show the effect of different thermal electron densities on the
cosmic ray electron population. The dashed line is the product of the inputs
in Model 3; the solid line is the result using the inputs in Model 10. The
electron densities in this model are an order of magnitude smaller than those
in Model 3. As can be seen, the cosmic ray electron population is not greatly
sensitive to changes in the thermal electron density in the higher energy range.

  In Fig.9 we show the constraints imposed on the time of the initial event in
 our composite model. The input parameters are from Model 3. The dashed line
corresponds to an evolutionary time of 1 Gyr, and the solid line is the
electron population after 1.4 Gyr.

  We also investigated the effect of a 5 $\mu$G field with regards to the
hypothesis that both the EUV excess and the high energy X-ray flux are produced
 by the ICS mechanism with the same population of electrons. In this work we
used Models 3, 8, 9, 10, and 11 which are capable of reproducing both the EUV
and synchrotron radio flux. We investigated whether any of these models were
also capable of producing the high energy X-ray flux by the ICS mechanism. The
results are shown in Fig.10. In this figure, the top line is the observed
excess hard energy flux (Rephaeli et al. 1999). The lower lines are the fluxes
produced by the ICS mechanism in all five models. The flux produced by the ICS
mechanism in all of these models is more than two orders of magnitude below the
 observed flux. Changes in the input characteristics of the injection events in
these different models do not affect the predicted flux significantly.

\section{DISCUSSION AND CONCLUSIONS}

  Atoyan \& V\"olk (2000) are the only authors who have explored the production
 of an EUV excess in a high ($\geq 1 \mu$G) intracluster field. They explored a
 model which is different in formal concept from those considered here, but is
 similar in form to our model with a large primary event followed by smaller
 injections. In specific, they considered two populations of cosmic rays. One is
 an old, highly evolved population which has undergone re-acceleration and
 compression; this population produces the EUV emission. A second higher
 energy population of cosmic rays which has been produced more recently by
 secondary origin is invoked to produce the radio emission. While this model
 may be appropriate, we note that models invoking secondary electrons for
 the radio emission have not been successful in producing the pronounced
 steeping of the spectral index of the radio emission with increasing distance
 from the cluster center that has been observed in well-studied radio halos
 such as Coma and Perseus. It has not been demonstrated that the Atoyan \&
 V\"olk model can surmount this obstacle.

  In this work we have studied general classes of models to establish which
 of these are capable of producing both the EUV and the synchrotron radio
emission in a $5 \mu$G cluster field. We studied models with a single initial
event, models with continuous injection events, and combinations of these two.
We find that in all models with just a single initial event, either
the EUV emission will be underproduced or the synchrotron radio emission will
be overproduced depending on the size of the event. With the passage of time,
synchrotron losses (which are dominant in the high field case) will remove the
higher energy components of the initial electron population, but the
distribution required to produce the synchrotron radio emission is not
produced. This problem is basic to all models with a single electron
injection event. Continuous models encounter a different problem. An electron
spectrum can be constructed which will produce the synchrotron radio emission.
However, this population is not capable of producing the EUV flux.

  The combination of a primary event followed by smaller continuous events can
produce the required electron distribution. However, the initial event must
have occurred at a time no earlier than 1.4 Gyr in the past or the higher
energy electrons will have been lost and insufficient EUV flux will be
 produced. The electron energy loss rate is sufficiently steep that even
substantially increasing the number of electrons in the initial event will not
substantially change this time constraint. On the other hand, the time of the
initial event must be no more recent than 1 Gyr ago or the synchrotron radio
emission will be overproduced. This requires that the injection event must have
 occurred in the time span 1.4 Gyr $<$ injection $<$ 1 Gyr.

  Finally, we investigated the effects of a 5$\mu$G field if both the EUV
excess and the high energy X-ray flux are produced by ICS mechanism with the
same population of electrons. The results shown in Figure 10 show that the high
 energy X-ray emission produced via the ICS process falls far short of the
observed flux for all of these models. We conclude it is impossible to produce
both the EUV flux and the hard X-ray emission by ICS from the same population
of electrons with reasonable electron injection scenarios. We note that Atoyan \& V\"olk (2000) have reached similar conclusions.

\acknowledgments

  The project is supported by the National Science Council of Taiwan (grant
number NSC 90-2112-M-008-040) and the Graduate Institute of Astronomy, NCU.
S.B. thanks the faculty and staff of the Graduate Institute of Astronomy, NCU,
and the National Center for Theoretical Science of Taiwan for their support
and warm hospitality during an extended stay.

\newpage
\begin{deluxetable}{lccccccc}
\tablecaption{}
\tablehead{
\colhead{Model} & \colhead{$N_o$} & \colhead{$p$} & 
\colhead{EI(erg)} & \colhead{$Q_o \,({\rm s}^{-1})$}
 &\colhead{$p'$} & \colhead{EI(erg yr$^{-1}$)}& 
\colhead{$n_e$(${\rm cm}^{-3}$)}\\
\cline{1-8} \\   
\multicolumn{8}{c}{Single Primary Event}} 
\startdata
  No.1 & $5.5\times{10}^{67}$ & 2.32 & $1.64\times{10}^{61}$ &   &   &  & 
$3\times {10}^{-3}$\\
\cutinhead{Continuous Injection}
  No.2 & $2\times {10}^{49}  $ & 2.32 & $5.96\times {10}^{42}$ & $2\times {10^{49}}$ & 2.32 &$1.88\times {10}^{50}$ & $3\times {10}^{-3}$\\

\cutinhead{Primary Event Followed by Continuous Injection}
  No.3 & $9\times {10}^{67}$ & 2.32 & $2.68\times {10}^{61}$& $2\times {10}^{49}$ & 2.32 & $1.88\times {10}^{50}$ & $3\times {10}^{-3}$  \\
  No.4 & $9\times {10}^{68}$ & 2.32 & $2.68\times {10}^{62}$& $2\times {10}^{49}$ & 2.32 & $1.88\times {10}^{50}$ & $3\times {10}^{-3}$  \\
  No.5 & $9\times {10}^{67}$ & 2.5  & $6.87\times {10}^{60}$& $2\times {10}^{49}$ & 2.32 & $1.88\times {10}^{50}$ & $3\times {10}^{-3}$  \\
  No.6 & $9\times {10}^{67}$ & 2.32 & $2.68\times {10}^{61}$& $2\times {10}^{49}$ & 2.2  & $4.80\times {10}^{50}$ & $3\times {10}^{-3}$ \\
  No.7 & $9\times {10}^{67}$ & 2.32 & $2.68\times {10}^{61}$& $2\times {10}^{48}$ & 2.32 & $1.88\times {10}^{49}$ & $3\times {10}^{-3}$ \\
  No.8 & $9\times {10}^{67}$ & 2.32 & $2.68\times {10}^{61}$& $8\times {10}^{48}$ & 2.2 & $1.92\times {10}^{50}$ & $3\times {10}^{-3}$ \\
  No.9 & $9\times {10}^{67}$ & 2.32 & $2.68\times {10}^{61}$& $9\times {10}^{49}$ & 2.5 & $2.17\times {10}^{50}$ & $3\times {10}^{-3}$ \\
  No.10& $9\times {10}^{67}$ & 2.32 & $2.68\times {10}^{61}$& $2\times {10}^{49}$ & 2.32 & $1.88\times {10}^{50}$ & $3\times {10}^{-4}$ \\
  No.11& $3.5\times {10}^{68}$ & 2.5 & $2.67\times {10}^{61}$& $2\times {10}^{49}$ & 2.32 & $1.88\times {10}^{50}$ & $3\times {10}^{-3}$ \\
\enddata
\end{deluxetable}

\newpage
\begin{figure} %1
\plotone{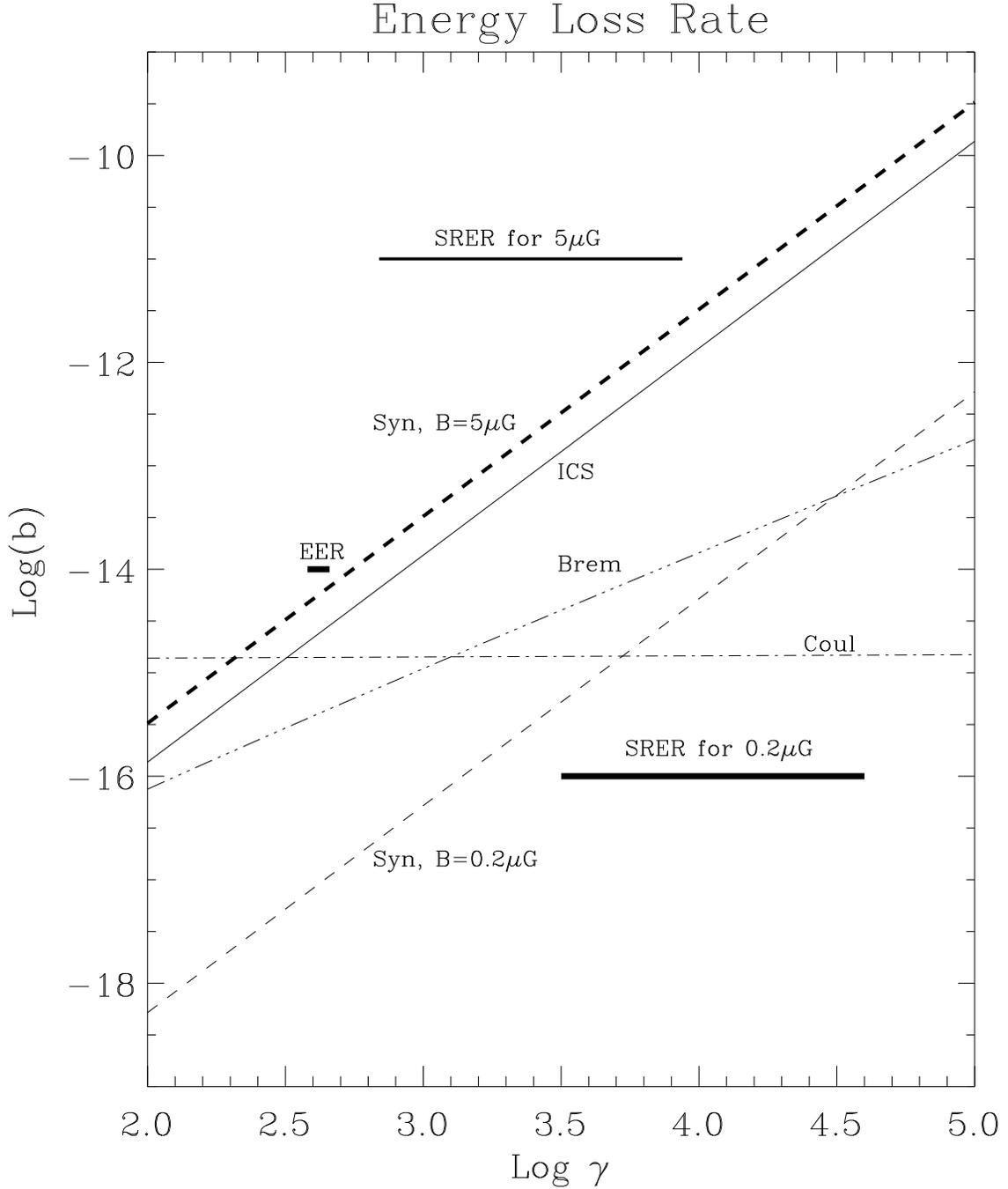}
\caption{Energy loss rates for individual mechanisms including inverse Compton
scattering (ICS), synchrotron radiation (Syn) in two magnetic
fields, Coulomb ionization (Coul), and bremsstrahlung radiation (Brem). EER is
the ``EUV-emitting electron range'' and SRER is the ``synchrotron-radio-emitting electron range''.}
\end{figure}
%----------------------
\newpage
\begin{figure} %2
\plotone{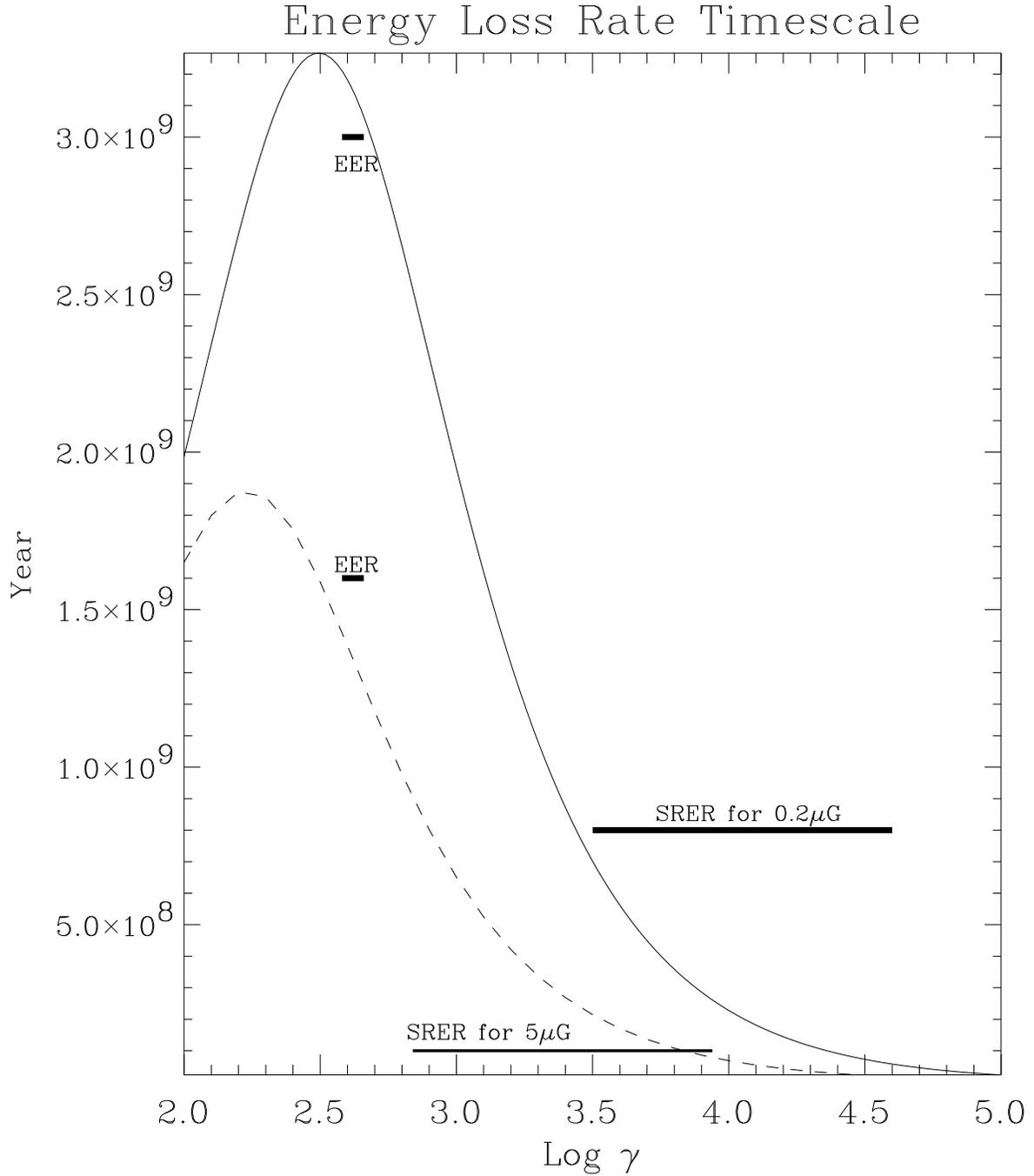}
\caption{The energy loss rate timescale for the electrons in a $5\mu$G and
$0.2\mu$G field. The dashed line is the timescale for $5\mu$G case and
the solid line is for $0.2\mu$G. The thick horizontal lines mark the EUV-emitting electron range and the synchrotron-radio-emitting electron range.}
\end{figure}

\newpage
\begin{figure} %3
\plotone{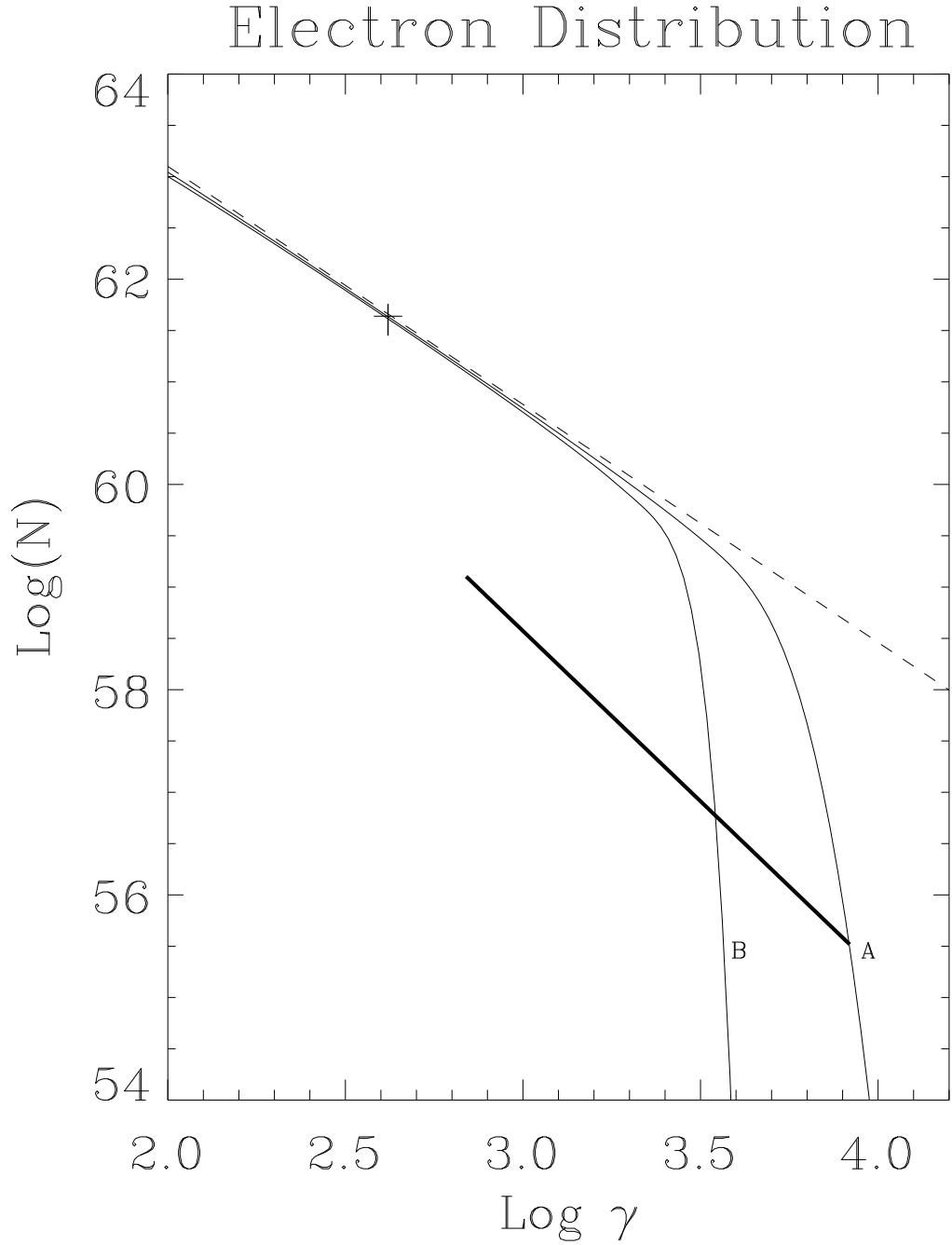}
\caption{The electron spectrum using the inputs in Model 1. The dashed line
 shows the distribution of the initial injected electrons. Line A shows the
distribution of the electrons after $1.4 \times {10}^8$ yr 
while line B shows the
distribution after $2.5 \times {10}^8$ yr. The cross is the distribution of the
electrons producing the observed EUV flux and the heavy solid line is the
distribution of the electrons producing the synchrotron radio emission.}
\end{figure}
%---------------------------------
\newpage
\begin{figure} %4
\plotone{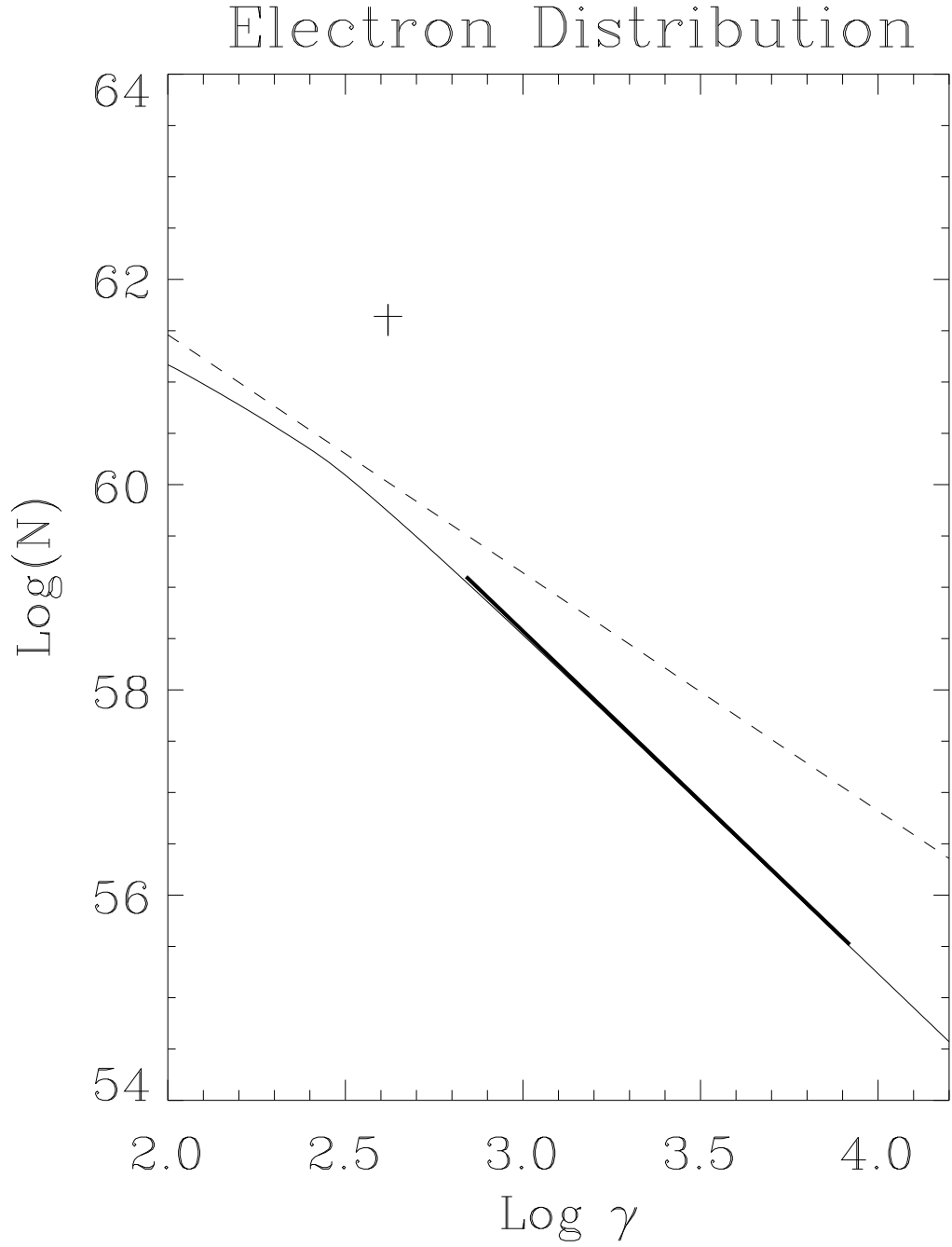}
\caption{The electron spectrum using the inputs in Model 2 with an evolution
time of $2 \times {10}^9$ yr. The dashed line shows the total injected
electrons.}
\end{figure}
%---------------------
\begin{figure} %5
\plotone{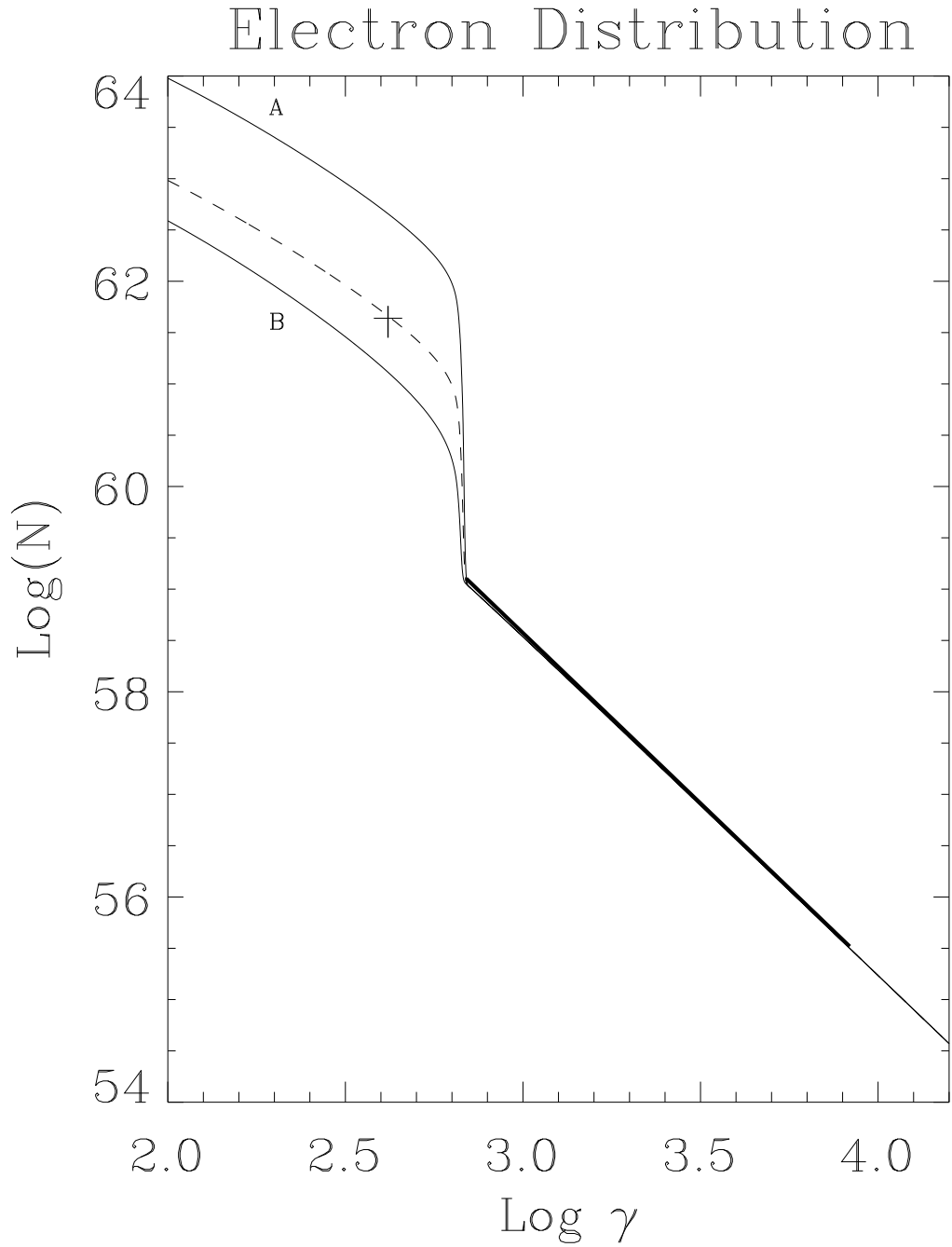}
\caption{The dashed line shows the electron distribution using inputs from
Model 3 with an evolution time of 1 Gyr. Line A with input parameters from
Model 4 shows the distribution of electrons with $N_o$ increased, and line B
with input parameters from Model 5 shows the distribution of the electrons
with $p$ increased.}
\end{figure}
%---------------------------
\newpage
\begin{figure} %6
\plotone{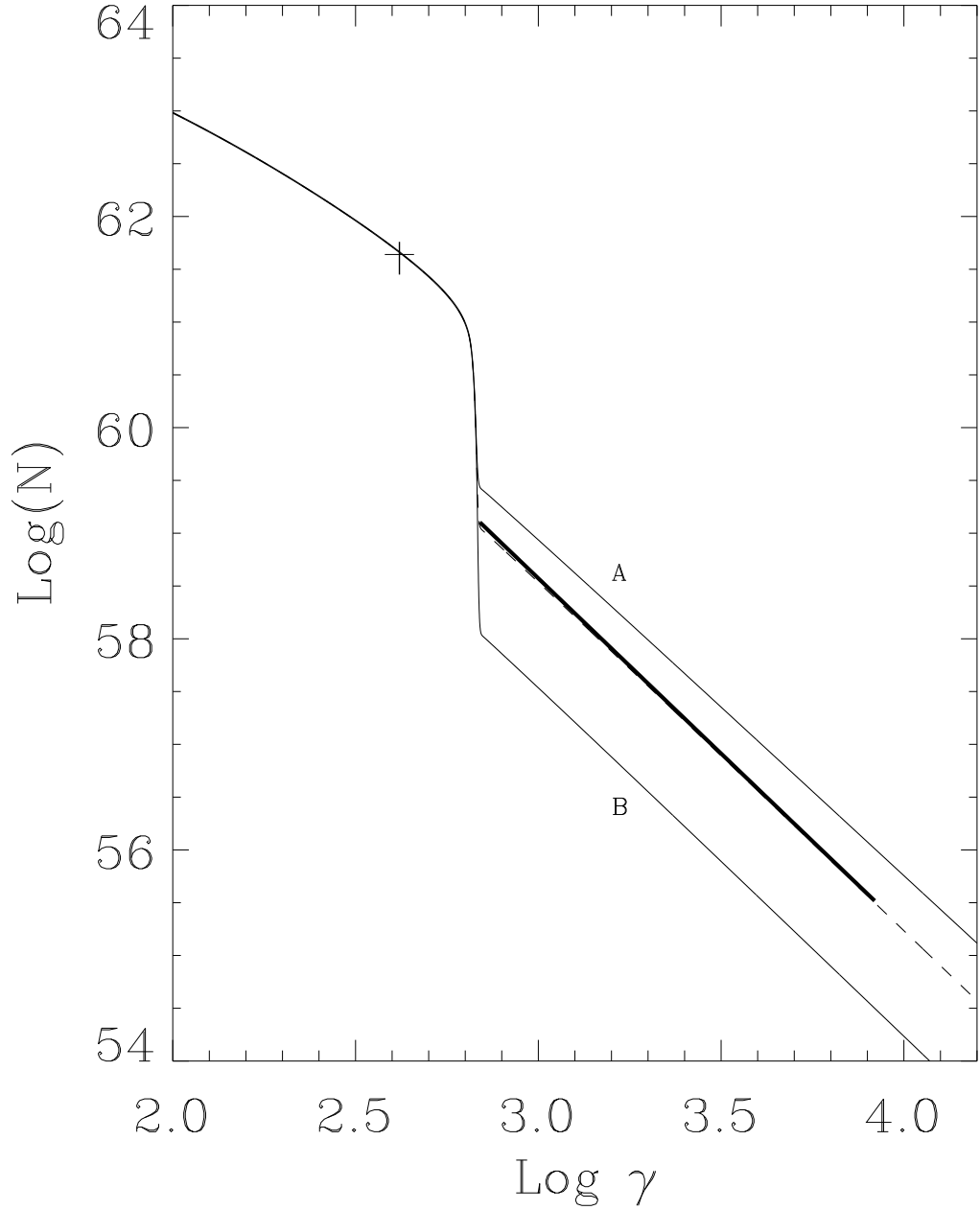}
\caption{The effects of changes in $Q_o$ and $p'$ on the electron spectrum. The
 dashed line is Model 3. Line A shows the results of decreasing $p'$ using the
input parameters of Model 6. Line B shows the results of decreasing $Q_o$ using
 the inputs in Model 7. The evolution time is 1 Gyr.}
\end{figure}
%------------------------------------------------
\newpage
\begin{figure} %7
\plotone{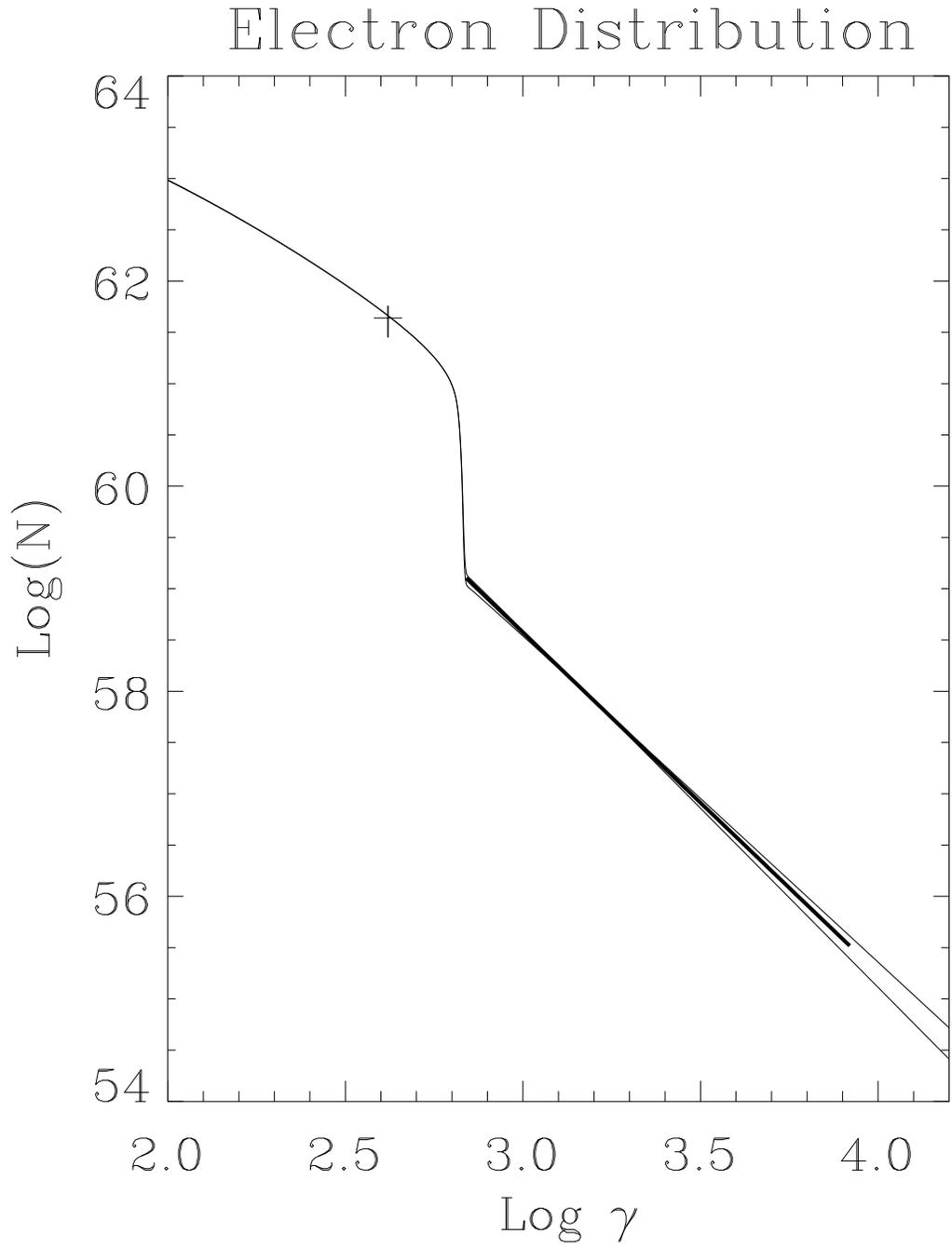}
\caption{Models mimic the best fit electron distribution (Model 3). 2.2 and 2.5
are chosen for $p'$ and the resultant electron distributions are shown as the
solid lines. The evolution time is 1 Gyr.}
\end{figure}
%--------------------------
\newpage
\begin{figure} %8
\plotone{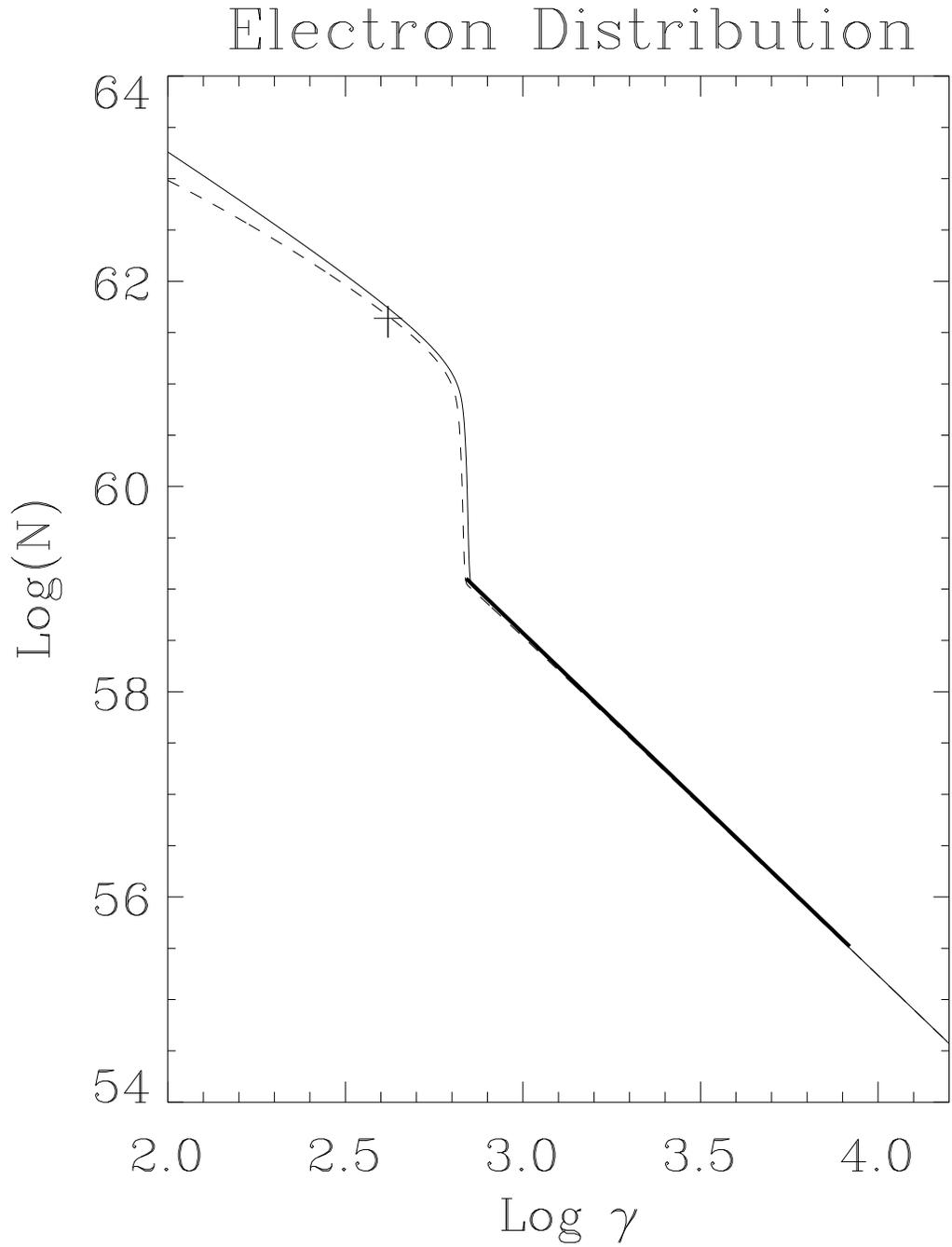}
\caption{The effect of differing thermal electron densities, $n_e$. The dashed
 line is the product of the inputs in Model 3; the solid line is the result
using the inputs from Model 8 with $n_e$ smaller by a factor of 10. The
evolution time is 1 Gyr.}
\end{figure}
%-----------------------------
\newpage
\begin{figure} %9
\plotone{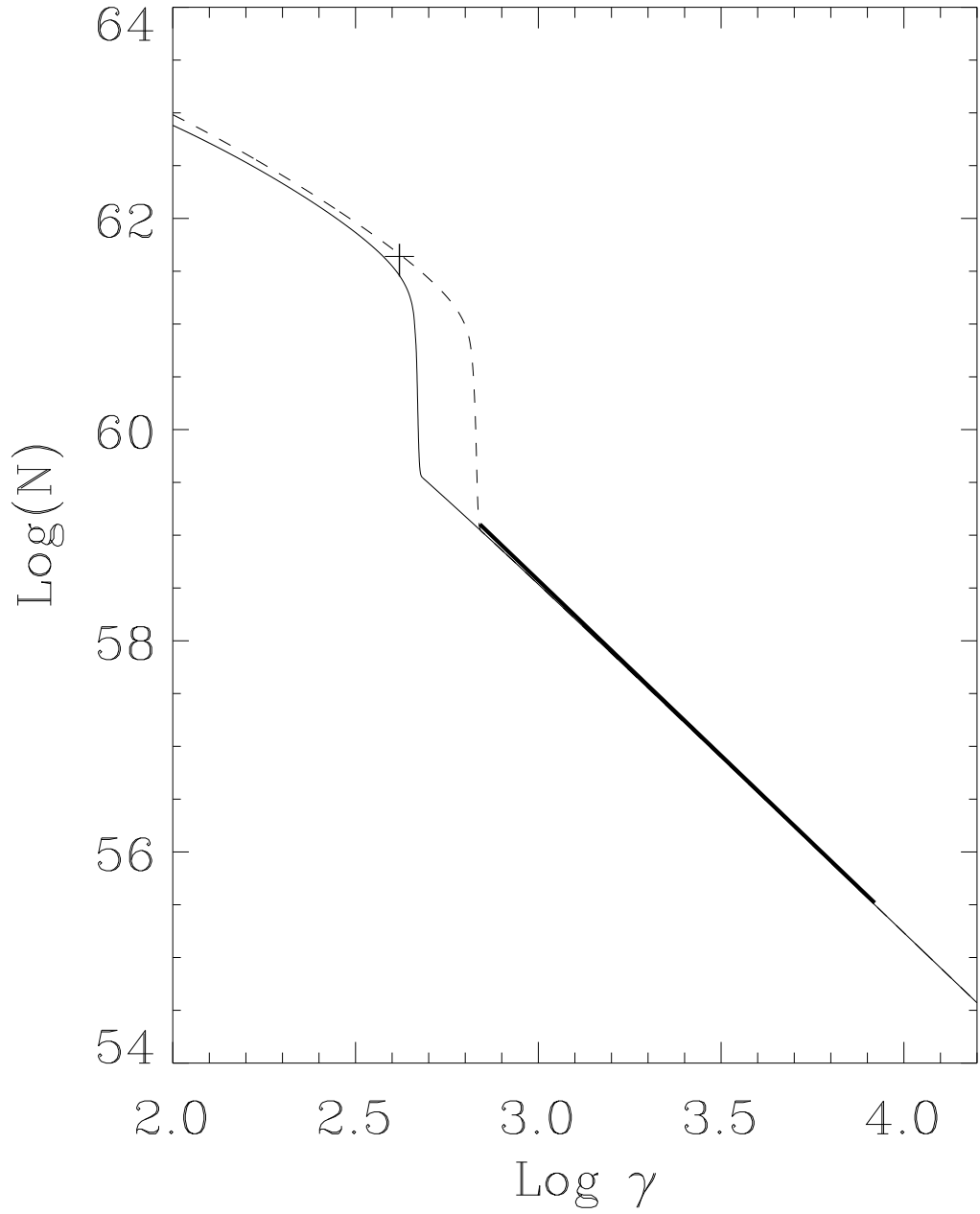}
\caption{The effect of different evolution times. The input parameters are from
 Model 3. The dashed line is the distribution after 1 Gyr and the solid line is
 the distribution after 1.4 Gyr.}
\end{figure}
%--------------------------
\newpage
\begin{figure} %10
\plotone{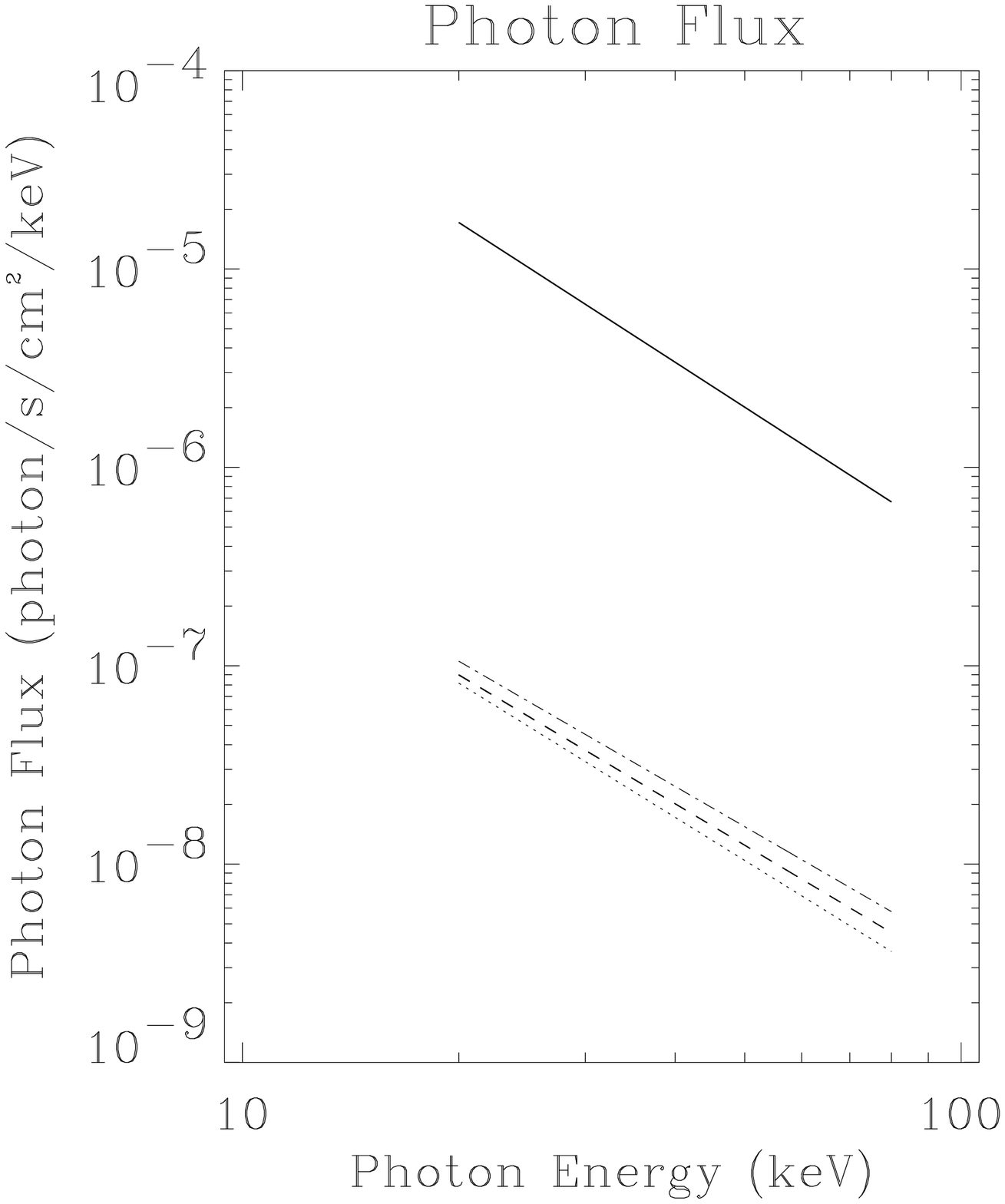}
\caption{A comparison of the observed hard X-ray flux and the ICS results of
the models. The solid line shows the excess high energy X-ray flux. The dashed
line shows the flux produced by the ICS mechanisms using inputs from Model 3,
10, and 11. The emission predicted by these three models is virtually the same.
The  dash-dot line shows the emission  predicted by Model 8,
and the dotted line shows the emission  predicted by Model 9.}
\end{figure}
%--------------------------

\begin{thebibliography}{}

\bibitem{1} Atoyan, A., \& V\"olk, H. 2000, \apj, 535, 45

\bibitem{2} Blumenthal, G. R., \& Gould, R. J. 1970, Rev. Mod. Phys.,
42, 237

\bibitem{3} Bonamente, M., Lieu, R., \& Mittaz, J. 2001a, \apj, 546, 805

\bibitem{4} Bonamente, M., Lieu, R., \& Mittaz, J. 2001b, \apj, 547, 7

\bibitem{5} Bowyer, S., \& Bergh\"ofer, T. 1998, \apj, 506, 502

\bibitem{6} Bowyer, S., Bergh\"ofer, T. W., \& Korpela, E. J. 1999, 
\apj, 526, 592

\bibitem{7} Brunetti, G., Setti, G., Feretti, L., \& Giovannini, G. 2001,
New Astronomy, 6, 1

\bibitem{8} Clarke, T., Kronberg, P., \& B\"ohringer, H. 2001, \apj,
547, L111

\bibitem{9} Deiss, B.M., Reich, W., Lesch, H., \& Wielebinski, R. 1997,
\aap, 321, 55

\bibitem{10} Dixon, W. V. D., Hurwitz, M., \& Ferguson, H. 1996, \apj, 469, L77

\bibitem{11} Dixon, W. V. D., Sallmen, S., Hurwitz, M., \& Lieu, R. 2001a,
\apj, 550, L25

\bibitem{12} Dixon, W., Sallmen, S., Hurwitz, M., \& Lieu, R. 2001b,
\apj, 552, L69

\bibitem{13} En$\beta$lin, T. A., \& Biermann, P. L. 1998, \aap, 330, 90

\bibitem{14} Feretti, L., Dallacasa, D., Giovannini, G., \& Tagliani, A.
1995, \aap, 302, 680

\bibitem{15} Fusco-Femiano, R., Dal Fiume, D., Feretti, L., Giovannini, G.,
Grandi, P., Matt, G., Molendi, S., \& Santangelo, A. 1999, \apj, 512, L21

\bibitem{16} Giovannini, G., Feretti, L., Venturi, T., Kim, K. -T.,
 \& Kronberg, P. P. 1993, \apj, 406, 399

\bibitem{17} Hwang, C.-Y. 1997, Science, 278, 1917

\bibitem{18} Kaastra, J. S., Ferrigno, C., Tamura, T., Paerels, F. B. S.,
Peterson, J. R., \& Mittaz, J. P. P. 2001, \aap, 365, L99

\bibitem{19} Lieu, R., Mittaz, J., Bowyer, S., Breen, J., Lockman, F.,
Murphy, E., \& Hwang, C.-Y. 1996, Science, 274, 1335

\bibitem{20} Lieu, R., Bonamente, M., \& Mittaz, J. 1999, \apj, 517, 91

\bibitem{21} Lieu, R., Bonamente, M., Mittaz, J., Durrett, F., Dos Santos,
S., \& Kaastra, J. 1999, \apj, 527, 77

\bibitem{22} Lieu, R., Bonamente, M., \& Mittaz, J. 2000, \aap, 364, 497

\bibitem{23} Longair, M. S. 1994, High Energy Astrophysics, Vol. 2,
Cambridge University Press

\bibitem{24} Mittaz, J., Lieu, R., \& Lockman, F. 1998, \apj, 498, 17

\bibitem{25} Peterson, J. R., Paerels, F. B. S., Kaastra, J. S., Arnaud, M.,
 Reiprich, T. H., Fabian, A. C., Mushotzky, R. F., Jernigan, J. G., \&
Sakelliou, I. 2001, \aap, 365, L104

\bibitem{26} Rephaeli, Y. 1988, Comments on Modern Physics, Part C, 12, 265

\bibitem{27} Rephaeli, Y., Gruber, D., \& Blanco, P. 1999, \apj, 511, L21

\bibitem{28} Sarazin, C. L. 1999, \apj, 520, 529

\bibitem{29} Sirk, M., Vallerga, J., Finley, D., Jalensky, P., \& Malina, R.
1997, \apj, 110, 347

\bibitem{30} Tamura, T., Kaastra, J. S., Peterson, J. R., Paerels, F. B. S.,
Mittaz, J. P. D., Trudolyubov, S. P., Stewart, G., Fabian, A. C.,
Mushotzky, R. F., Lumb, D. H., \& Ikebe, Y. 2001, \aap, 365, L87

\end{thebibliography}
\end{document}